# Impermeable Barrier Films and Protective Coatings Based on Reduced Graphene Oxide

Y. Su[1], V. G. Kravets[1], S. L. Wong[1], J. Waters[2], A. K. Geim[1], R.R. Nair[1]

[1]School of Physics and Astronomy, University of Manchester, Manchester M13 9PL, UK

[2]School of Earth, Atmospheric and Environmental Sciences, University of Manchester, Manchester M13 9PL, UK

*Barrier films preventing permeation of gases and moisures are important for many industries ranging from food to medical and from chemical to electronic. From this perspective, graphene has recently attracted particular interest because its defect free monolayers are impermeable to all gases and liquids. However, it has proved challenging to develop large-area defectless graphene films suitable for industrial use. Here we report barrier properties of multilayer graphitic films made by chemical reduction of easily and cheaply produced graphene oxide laminates. They are found to provide a practically perfect barrier that blocks all gases, liquids and aggressive chemicals including, for example, hydrofluoric acid. In particular, if graphene oxide laminates are reduced in hydroiodic acid, no permeation of hydrogen and water could be detected for films as thin as 30 nm, which remain optically transparent. The films thicker than 100 nm become completely impermeable. The exceptional barrier properties are attributed to a high degree of graphitization of the laminates and little structural damage during reduction. This work indicates a close prospect of thin protective coatings with stability and inertness similar to that of graphene and bulk graphite, which can be interesting for numerous applications.*

Membranes made from graphene and its chemical derivative called graphene oxide[1-3] (GO) show a range of unique barrier properties[4-6]. Defect-free monolayer graphene is impermeable to all gases and liquids[4] and, similar to graphite, shows high chemical and thermal stability with little toxicity. These characteristics are believed to provide graphene with a competitive edge over the existing barrier materials[7]. Unfortunately, prospects of using graphene as a protective coating have so far been hampered by difficulties of growing large area defect-free films. For example, it is shown that graphene films grown by chemical vapor deposition (CVD) possess many defects and grain boundaries and do not protect copper against oxidation but, to the contrary, speed up its corrosion[8]. A potential solution to this problem is the use of graphene-based multilayers[9-11]. In this respect, GO is particularly attractive because multilayer films can be produced easily and relatively cheaply by depositing GO solutions onto various substrates by spraying, dip coating, etc. The resulting GO laminates are shown to exhibit highly unusual permeation properties[5,6]. In the dry state, they are completely impermeable but, in humid conditions, provide no barrier for water vapour[5]. If immersed in water, the laminates act as molecular sieves allowing transport of small ions and blocking large ones[6]. Although such unique and contrasting properties may be useful for certain applications, barrier films are generally required to have ultra-low water permeation rates of <$10^{-6}$ g/m$^2$ per day under ambient conditions[7]. One of the possible strategies to overcome the remaining issue of water permeation is to reduce GO back to graphene.

Considerable efforts have recently been made to utilize multilayer graphene-based films (thermally-reduced GO (T-RGO)[9,10], CVD graphene[11] and graphene-based composites[7]) as ultra-barriers for organic electronics and as oxidation resistant and anticorrosion coatings[7,12-18]. However, T-RGO membranes are extremely fragile and contain many structural defects, which results in notable water permeation[5,7]. A high density of defects in CVD graphene limits its possible uses too[7,8]. Similarly, GO-polymer composites have so far exhibited gas permeability too high to consider them for realistic applications[7]. To increase the quality of GO-based coatings, it is essential to decrease the number of defects formed during the reduction process[19-24]. Recent studies show that the use of hydroiodic (HI) acid as a reducing agent results in RGO's quality being much better than what other reduction techniques could provide in terms of electrical and mechanical properties[21,23,25]. The HI reduction leaves fewer structural defects and little deformation so that the mechanical strength increases becoming even higher than that of initial GO laminates which are known to be already exceptionally strong[1]. This additional strengthening could be due to chemical crosslinking of graphene sheets by remnant functional groups and/or iodine[21]. Another interesting reducing agent is ascorbic acid, that is, vitamin C (VC)[22,24]. It shows not only good reducing characteristics but stands out as environmentally friendly and nontoxic, which may be a critical factor in many applications. Accordingly, our work has focused on barrier properties of GO laminated films obtained by thermal reduction and exposure to HI and VC.

Graphite oxide was prepared by the Hummers method[26] and then dispersed in water by sonication, which resulted in stable GO solutions[1-3]. The size of individual GO flakes varied from approximately $0.2 \times 0.2$ $\mu m^2$ to $20 \times 20$ $\mu m^2$, and we did not find any notable size dependence in the barrier properties described below. We used two types of RGO samples: free-standing membranes and supported thin films made on various substrates. The freestanding membranes were fabricated by vacuum filtration as described in ref. 5 and had thickness $d$ from 0.5 to 5 $\mu m$. The supported films were prepared by rod-coating[27] or spray-coating on top of a polyethylene terephthalate (PET) film (12 $\mu m$ thick), metal foils and oxidized silicon wafers. Thermal reduction was carried out at 300 $^0C$ for 4 hours in a hydrogen-argon mixture. HI reduction[21,23] was done by exposing GO films to the acid vapor at 90 $^0C$. The exposure time varied from 5 to 30 minutes, depending on $d$. Then, samples were repeatedly rinsed with ethanol to remove residual HI. For VC reduction[22,24], GO films were immersed in water solution of VC (30 g/l) for 1 hour at 90 $^0C$. Permeation properties of the resulting RGO films were measured using several techniques that were described in detail previously[5,6]. In brief, for vapor permeation measurements, free-standing membranes and RGO films on PET were glued to a Cu foil with an opening of 2 cm in diameter. The foil was clamped between two O-rings sealing a metal container. Permeability was measured by monitoring the weight loss of the container that was filled with water and other liquids inside a glovebox[5]. In gas experiments[5], GO-on-PET films were placed between two rubber gaskets and pressurized from one side to up to 1 bar. Gas permeation was monitored on the opposite (vacuum) side by using mass spectrometry. We used INFICON UL200 that allowed detection of helium and hydrogen.

Figure 1a shows examples of water vapor permeation measurements through GO and RGO membranes. In agreement with the previous reports[5,6], non-reduced membranes are completely impermeable to all gases, except for water that evaporates with little resistance. After thermal reduction, the same membrane exhibits 10 times slower water loss but becomes fragile[5]. Permeation rates for different T-RGO samples are found to vary by a factor of 2. In contrast, GO membranes reduced in VC (VC-RGO) exhibit highly reproducible behavior and a decrease by 5 orders of magnitude in water permeability with respect to pristine GO laminates (Fig. 1b). HI-reduced GO membranes (HI-RGO) provide even a better barrier such that water permeation is undetectable within our best accuracy of ≈0.1 mg per week. This yields an upper limit for water permeation through HI-RGO films of submicron thickness as ~$10^{-11}$ mm×g/cm$^2$×s×bar (see Fig. 1b), which is two orders of magnitude lower than water transparency for the industry-standard barrier films (aluminized PET)[28,29]. We also performed permeation measurements for several organic solvents such as acetone, methanol, ethanol and propanol. No permeation could be detected for either original GO (as reported previously[5]) or reduced membranes.

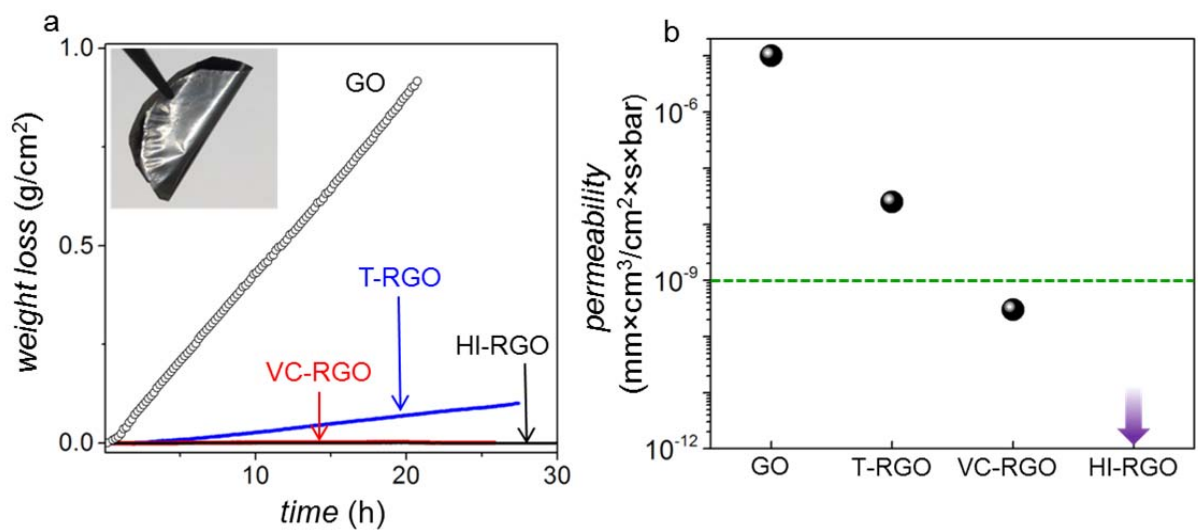

*Fig. 1. Water permeation through free standing multilayer graphene membranes. a – Water loss from a container sealed with GO and RGO membranes with a diameter of ≈2 cm (d ≈0.5 μm). For GO, the loss is the same as through an open aperture and limited by water evaporation. Inset: photo of an HI-RGO membrane (diameter ≈2 cm). b – Permeability of various RGO membranes with respect to water vapor. The arrow indicates our detection limit. Permeability of pristine GO is taken from ref. 5. Green line: water permeability for the industrial-standard barrier films (Al on PET)[28,29].*

It is more practical to use RGO not as free-standing membranes but as thin coatings on top of other materials. To evaluate barrier properties of such coatings, we have employed standard PET films (12 μm thick) as a support for RGO. Figure 2 shows an optical photo of the PET film covered with 30 nm of HI-RGO (see Supplementary S1). Within our experimental accuracy, such barrier films show no detectable permeation of either hydrogen or water (Fig. 2b). However, they still remain slightly permeable to He, which can be attributed to occasional microscopic pinholes in such thin laminates that are made of randomly-stacked graphene crystallites[5,6]. It requires RGO films thicker than 100 nm to block permeation of He completely, beyond He-leak detection that is the most sensitive method to check for potential leaks in high vacuum equipment (upper inset of Fig. 2a). We have also tested a VC-RGO coating on PET and didn't find any significant difference with respect to HI-RGO. Importantly, chemically reduced GO exhibits strong adhesion to PET and, despite sub-μm thickness, withstands folding and moderate scratching which allows normal handling procedures similar to aluminized PET.

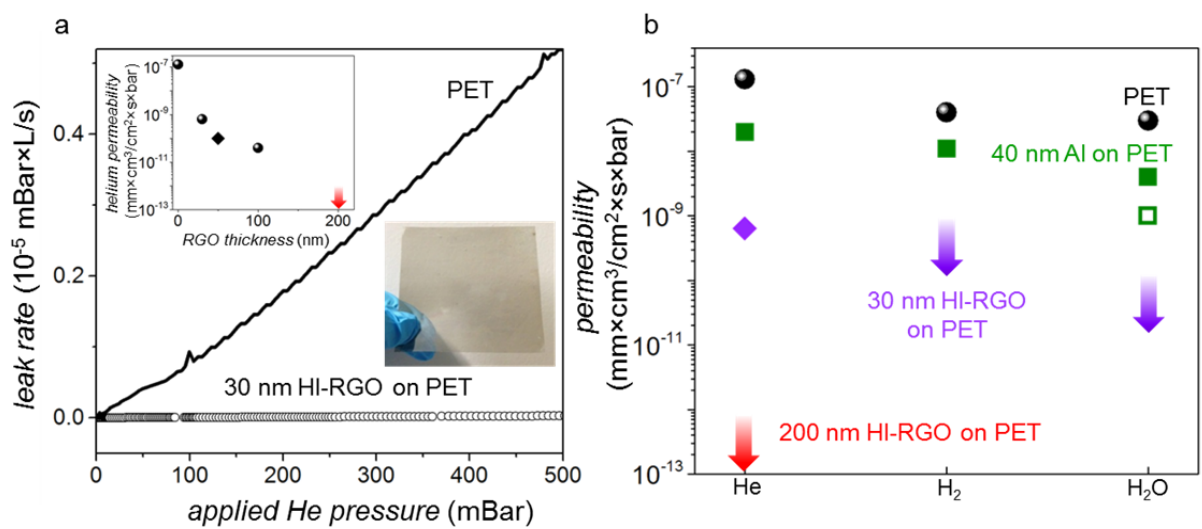

*Fig. 2. Permeation through RGO barrier coatings. a* – He-leak measurements for a bare PET film (12 μm) and the same film coated with 30 nm of HI-RGO. The latter film is shown in the lower inset and exhibits an optical transparency of ≈35%. The transparency is reduced to 7% for 100 nm thick RGO, and coatings thicker than 200 nm are opaque. Upper inset: He-permeability as a function of HI-RGO thickness (circles). The diamond symbol is for VC-RGO coating (d ≈50 nm). *b* – Barrier properties of bare PET, HI-RGO on PET and aluminized PET with respect to He, $H_2$ and $H_2O$. Bare films show permeability in agreement with literature values[30]. The green symbols are for a 40 nm thick aluminum film on PET (solid squares, our measurements; open symbol, the literature value[28,29]). The violet and read arrows indicate our detection limits for 30 and 200 nm HI-RGO, respectively.

The observed superior barrier properties of RGO with respect to gases and liquids suggest possible uses of RGO as anticorrosion and chemical-resistant coatings[11,13,17]. To evaluate their barrier properties with respect to salts and acids, we have used the measurement setup described in ref. 5. Briefly, two compartments of a U-shaped container were separated by a free-standing membrane and then filled with pure water and a 1M salt solution. Salt diffusion was monitored by measuring its concentration in the pure water compartment by using ion chromatography and gravimetric analysis[5]. Figure 3a illustrates our results for the case of NaCl. Cl ions permeate rapidly through non-reduced GO membranes, in agreement with the earlier report[6]. However, within our experimental accuracy[6], no salt permeation could be detected through either HI- or VC- RGO membranes.

To further illustrate anticorrosion properties of our graphitic laminates, we have tested their protection against HF, one of the most corrosive acids. The upper insets of Fig. 3a show the effect of HF on oxidized Si wafers (300 nm of $SiO_2$), which were protected with 0.5 μm thick films of GO and VC-RGO. A drop of concentrated HF was placed on top of the coatings and continuously maintained for several hours. Then the coating was peeled off to assess damages. As evident from Fig. 3a, HF permeated through the GO film, as expected[6], and etched through the entire thickness of the $SiO_2$ layer. On the other hand, the RGO film fully protected the wafer against HF, and no sign of $SiO_2$ etching could be detected within our accuracy of better than 10 nm. Similar acid drop tests were carried out for RGO coatings on top of various metals, including Cu and Ni. The latter foils were exposed to nitric and hydrochloric acids in different concentrations (0.1 to 10 M) but no degradation of the surface could be observed after several days of exposure. Furthermore, VC-RGO coated Ni and steel foils were immersed in saturated iron chloride and sodium chloride solutions for many days and, again, no degradation could be detected. Finally, we covered glass petri dishes with HI-RGO (bottom inset of Fig. 3a). This graphitic lining allowed the glassware to use HF.

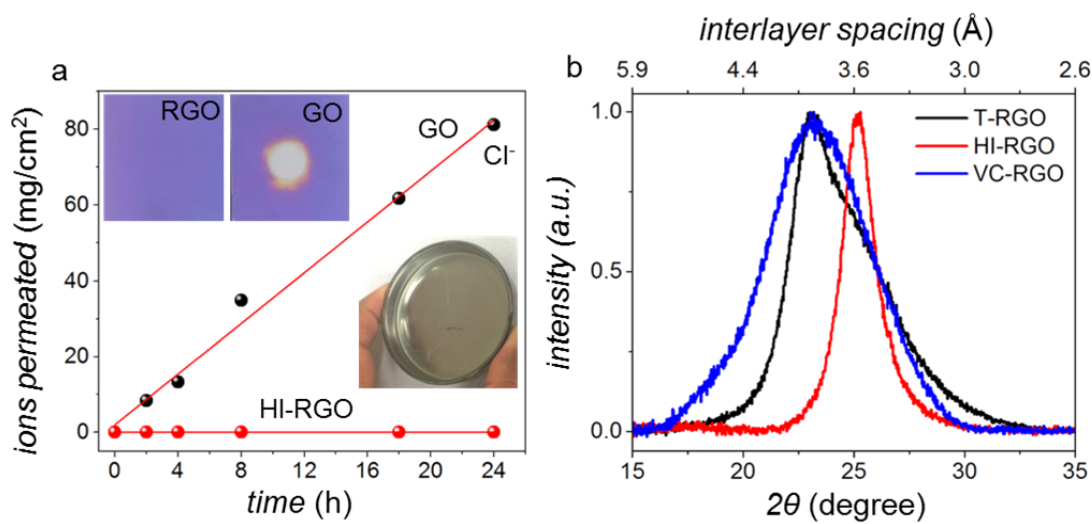

*Fig. 3. Chemical protection by RGO. a* – *Measurements of Cl⁻ permeation through GO and HI-RGO membranes (d ≈1 μm). Upper inset: Right and left photos (1cm × 1cm) show the effect of HF on oxidized Si wafers protected by GO and RGO, respectively. $SiO_2$ is completely removed in the white center region. Etching away of just 10 nm of $SiO_2$ would be visible as changes in the interference color, which are absent in the left image. Bottom inset: Glass petri dish lined with HI-RGO (≈1 μm thick).* ***b*** *– X-ray diffraction for thermally, HI- and VC- reduced GO membranes.*

Finally, we note that most substrates used in our work had smooth surfaces (PET, metal foils). However, we have also tried GO coating of materials with rough surfaces such as, for example, conventional bricks. Despite brick's highly porous structure, VC-RGO films exhibited excellent barrier properties in this case (Supplementary S2). Furthermore, adhesion of RGO films to metal surfaces is found to be weaker than to plastic ones. This makes the protective coatings of metals more prone to mechanical scratching and peeling off. To overcome this problem, we have mixed a GO solution with polyvinyl alcohol (PVA) and used the binary solution to make GO-PVA films in the same manner as described above. The resulting dry composites contained 30±10% of PVA. After their chemical reduction, the coatings exhibited nearly the same barrier properties as without PVA but with much improved adhesion and mechanical characteristics (see Supplementary S3).

To explain the observed barrier properties of RGO, we recall that permeation through non-reduced GO laminates occurs via a network of graphene capillaries filled with one or two monolayers of water[5,6]. The capillaries have a width varying from 0.7 nm to 1.3 nm, depending on humidity. After chemical or thermal reduction, these capillaries collapse, and the interlayer separation decreases to only ≈0.36 nm, which is close to the interlayer separation in graphite[3] (see Fig. 3b). This means that there is no space left for helium, water and other molecules to permeate between graphene sheets, and the only diffusion path remaining after the reduction is through structural defects. The crystallographic quality of reduced laminates can generally be judged by their X-ray diffraction peaks. Fig. 3b shows that HI-RGO exhibits the sharpest peak indicating the highest degree of graphitization[3,23]. VC-RGO has a broader X peak (Fig. 3b) but nonetheless shows barrier properties similar to those of HI-RGO. The only noticed difference between HI- and VC- RGO membranes was in their barrier properties with respect to water vapor (see Fig. 1b). The remnant $H_2O$ leakage for VC-RGO can be attributed to difficulties in reducing the membranes over their entire thickness by using VC that has larger and less mobile molecules than HI. However, approximately the same quality of graphitization for VC- and T- RGO films (Fig. 3b) indicates that factors other than the interlayer distance are important. We believe that the critical difference lies in the amount of structural defects formed during the reduction process. Indeed, it is known that during thermal reduction of GO, oxygen containing functional groups are removed together with carbon atoms from graphene planes, which results in release of CO and $CO_2$ gases[3,19]. They have to escape from the interior and, therefore, can delaminate and damage the layered structure. On the contrary, chemical reduction by using HI and VC is much gentler, and the functional groups attached to graphene sheets react with the reducing agents releasing water instead of gases[3,19], which can move along capillaries until they completely close[5,6]. As a result, chemically reduced GO is less damaged and retains a better structural order, which is also apparent from its shiny appearance whereas T-RGO looks lusterless.

In conclusion, chemically reduced GO films (especially, using HI) exhibit practically ideal barrier properties with respect to all tested gases, liquids, salts and acids. The films can be considered as thin graphitic linings, which can be produced on an industrial scale by simple solution processing. Taking into account that graphite is one of the most stable and chemically inert materials, this work opens a venue for many applications in which a barrier against moisture and oxygen is required and for the use in chemical and corrosion protection. The possibility to use the environmentally friendly reduction in ascorbic acid widens the scope of possible applications to sensitive areas including pharmaceuticals.

References
1. Zhu, Y. *et al.* Graphene and graphene oxide: synthesis, properties, and applications. *Adv. Mater.* **22**, 3906-3924 (2010).
2. Loh, K. P., Bao, Q., Eda, G. & Chhowalla, M. Graphene oxide as a chemically tunable platform for optical applications. *Nature Chem.* **2**, 1015-1024 (2010).
3. Pei, S. & Cheng, H.-M. The reduction of graphene oxide. *Carbon* **50**, 3210-3228 (2012).
4. Bunch, J. S. *et al.* Impermeable atomic membranes from graphene sheets. *Nano Letters* **8**, 2458-2462 (2008).
5. Nair, R. R., Wu, H. A., Jayaram, P. N., Grigorieva, I. V. & Geim, A. K. Unimpeded permeation of water through helium-leak-tight graphene-based membranes. *Science* **335**, 442-444 (2012).
6. Joshi, R. K. *et al.* Precise and Ultrafast molecular sieving through graphene oxide membranes. *Science* **343**, 752-754 (2014).
7. Yoo, B. M., Shin, H. J., Yoon, H. W. & Park, H. B. Graphene and graphene oxide and their uses in barrier polymers. *J. Appl. Polymer Sci.* **131**, 39628 (2014).
8. Schriver, M. *et al.* Graphene as a long-term metal oxidation barrier: worse than nothing. *ACS Nano* **7**, 5763-5768 (2013).
9. Kang, D. *et al.* Oxidation resistance of iron and copper foils coated with reduced graphene oxide multilayers. *ACS Nano* **6**, 7763-7769 (2012).
10. Yamaguchi, H. *et al.* Reduced graphene oxide thin films as ultrabarriers for organic electronics. *Adv. Energy Mater.* **4**, 1300986 (2014).
11. Prasai, D., Tuberquia, J. C., Harl, R. R., Jennings, G. K. & Bolotin, K. I. Graphene: corrosion-inhibiting coating. *ACS Nano* **6**, 1102-1108 (2012).
12. Huang, H.-D. *et al.* High barrier graphene oxide nanosheet/poly(vinyl alcohol) nanocomposite films. *J. Membrane Sci.* **409-410**, 156-163 (2012).

13  Kirkland, N. T., Schiller, T., Medhekar, N. & Birbilis, N. Exploring graphene as a corrosion protection barrier. *Corrosion Sci.* **56**, 1-4 (2012).
14  Yang, J. *et al.* Thermal reduced graphene based poly(ethylene vinyl alcohol) nanocomposites: enhanced mechanical properties, gas barrier, water resistance, and thermal stability. *Indust. Engineer. Chem. Res.* **52**, 16745-16754 (2013).
15  Yang, Y. H., Bolling, L., Priolo, M. A. & Grunlan, J. C. Super gas barrier and selectivity of graphene oxide-polymer multilayer thin films. *Adv. Mater.* **25**, 503-508 (2013).
16  Tseng, I. H., Liao, Y.-F., Chiang, J.-C. & Tsai, M.-H. Transparent polyimide/graphene oxide nanocomposite with improved moisture barrier property. *Mater. Chem. Phys.* **136**, 247-253 (2012).
17  Nilsson, L. *et al.* Graphene coatings: probing the limits of the one atom thick protection layer. *ACS Nano* **6**, 10258-10266 (2012).
18  Guo, F. *et al.* Graphene-based environmental barriers. *Environ. Sci. Tech.* **46**, 7717-7724 (2012).
19  Chua, C. K. & Pumera, M. Chemical reduction of graphene oxide: a synthetic chemistry viewpoint. *Chem. Soc. Rev.* **43**, 291-312 (2014).
20  Some, S. *et al.* High-quality reduced graphene oxide by a dual-function chemical reduction and healing process. *Scientific Rep.* **3**, 1929 (2013).
21  Pei, S., Zhao, J., Du, J., Ren, W. & Cheng, H.-M. Direct reduction of graphene oxide films into highly conductive and flexible graphene films by hydrohalic acids. *Carbon* **48**, 4466-4474 (2010).
22  Zhang, J. *et al.* Reduction of graphene oxide via L-ascorbic acid. *Chem. Commun.* **46**, 1112-1114, (2010).
23  Moon, I. K., Lee, J., Ruoff, R. S. & Lee, H. Reduced graphene oxide by chemical graphitization. *Nature Commun.* **1**, 73 (2010).
24  Fernández-Merino, M. J. *et al.* Vitamin C is an ideal substitute for hydrazine in the reduction of graphene oxide suspensions. *J. Phys. Chem. C* **114**, 6426-6432 (2010).
25  Su, Y., Du, J., Sun, D., Liu, C. & Cheng, H. Reduced graphene oxide with a highly restored π-conjugated structure for inkjet printing and its use in all-carbon transistors. *Nano Research* **6**, 842-852 (2013).
26  Hummers, W. S. & Offeman, R. E. Preparation of graphitic oxide. *J. American Chem. Soc.* **80**, 1339-1339 (1958).
27  Wang, J. *et al.* Rod-coating: towards large-area fabrication of uniform reduced graphene oxide films for flexible touch screens. *Adv. Mater.* **24**, 2874-2878 (2012).
28  Garnier, G. r., Yrieix, B., Brechet, Y. & Flandin, L. Influence of structural feature of aluminum coatings on mechanical and water barrier properties of metallized PET films. *J. Appl. Polymer Sci.* **115**, 3110-3119 (2010).
29  Lange, J. & Wyser, Y. Recent innovations in barrier technologies for plastic packaging - a review. *Packaging Tech. Sci.* **16**, 149-158 (2003).
30  Mercea, P. V. & Bârţan, M. The permeation of gases through a poly (ethylene terephthalate) membrane deposited with SiO2. *J. Membrane Sci.* **59**, 353-358 (1991).

# Supplementary Information

## S1. Optical and AFM characterisation of HI-RGO on PET

To characterize RGO films on PET, we have used scanning electron microscopy (SEM), atomic force microscopy (AFM) and optical absorption spectroscopy. Figure S1 shows an absorption spectrum for a 30 nm thick film of HI-RGO. For the visible spectrum the transmittance varies from ≈30 to 40%. The thickness of RGO coatings was measured using a Veeco Dimension 3100 AFM in the tapping mode under ambient conditions. The inset of Fig S1 shows a representative AFM image for a 30 nm thick HI-RGO on PET.

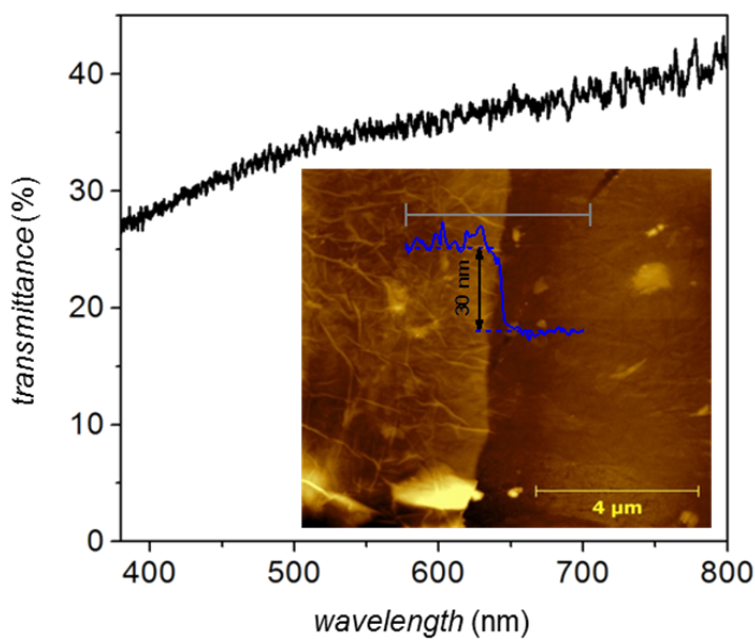

*Figure S1. Optical transmittance for a 30 nm thick HI-RGO on top of 12 μm thick PET with respect to bare PET. The inset shows an AFM image of the film near the boundary between bare PET and the RGO coating. Blue curve: Height profile along the gray line.*

## S2. RGO coating on rough and porous surfaces

To understand the effect of surface roughness and porosity on the barrier properties of our RGO films, we have deposited GO onto various surfaces. Those included polymer materials such as porous polycarbonate, polyvinylidene fluoride, polysulfone, etc. and extremely rough substrates such as brick and concrete surfaces. GO laminates on all these substrates were reduced by treating them with a VC solution at $80^0C$ for 2 hours or $50^0C$ for 24hours. We have found that, although the barrier quality can be sensitive to roughness and porosity, GO laminates provide a high permeation barrier for all tested surfaces. As an example, Figure S2 shows a photograph of a conventional red brick that is half coated with VC-RGO. If water is poured on the brick, it stays only on the part covered with highly hydrophobic RGO. One can quantify the barrier properties of VC-RGO by measuring the time required for disappearance of the water puddle (Fig. S2). The brick without any coating absorbs water within a few seconds. In contrast, water on top of the RGO coated part stays for many hours and eventually disappears mainly because of evaporation. Taking the evaporation into account, we estimate that VC-RGO treated bricks are ~4,000 times more water repellant than uncoated bricks.

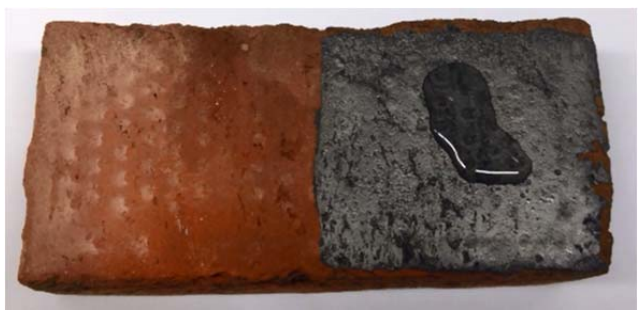

*Figure S2. Photograph demonstrating water permeation through a brick (~20 cm long) with and without VC-RGO coating. Brick without the graphitic coating rapidly absorbs water but it can stay on top of the VC-RGO coated part for many hours.*

**S3. Polyvinyl alcohol modified GO for improved adhesion**

Adhesion between treated surfaces and RGO is critical for the perspective use of such films as chemical and anticorrosion coatings. Adhesion of RGO to plastic and glass surfaces has been found strong. Qualitatively, the graphitic films were as robust as the standard barrier films (40 nm Al on PET) but the wear properties require further studies and quantification. In contrast, adhesion of RGO to metal surfaces was weak, which resulted in easy scratching and partial peeling of the protective coating. To overcome the drawbacks of weak adhesion to metal surfaces, we have employed the previous observation[1-3] that mechanical, electrical and biocompatible properties of GO laminates can be improved by interlayer cross-linking with PVA or other polymer molecules. For the purpose of this report, we have tested permeation properties of PVA-GO composite films, both before and after their chemical reduction.

PVA-GO samples were prepared by blending water solutions of GO and PVA by using a magnetic stirrer. The concentrations were chosen such that we achieved 60-80 weight percentage of GO in the final laminates, after evaporation of water. All the tested PVA-GO films exhibited similar properties, irrespective of their PVA fraction. We used vacuum filtration, drop casting and rod coating techniques to produce free standing PVA-GO membranes and PVA-GO coated substrates. Figure S3a shows examples of our permeation measurements for water and other organic vapors through a 1μm thick PVA-GO membrane, before and after its reduction with HI. Similar to GO, PVA-GO membranes completely block all gases and vapors except for water. After reduction of PVA-GO with HI, the water permeation is reduced approximately by four orders of magnitude (Fig. S3).

We have also studied salt permeation properties of such cross-linked GO membranes and found that permeation rates are beyond our detection limit, too. We have tested not only HI- but also VC- reduced PVA-GO and observed no major differences. The inset of Figure S3b shows an optical photograph of a steel plate coated with VC-reduced PVA-RGO. Such protecting coatings exhibit good adhesion to metal surfaces including copper, steel, nickel, etc. Copper foils coated with VC-reduced PVA-RGO were tested for acid corrosion. We could not detect any sign of corrosion in tests similar to those described in the main text and involving oxidized Si wafers protected with unmodified RGO (Fig. 3a of the main text).

Fig. S3b shows X-ray diffraction for HI-reduced PVA-RGO membranes. They exhibit a layered structure similar to HI-RGO but with an interlayer separation of ≈4.2 Å, that is, considerably larger than in the membranes without PVA (see Fig. 3b of the main text). This increase in the interlayer distance is attributed to the presence of PVA molecules between reduced GO sheets (intercalation-like composites). Although the interlayer distance increases, the presence of polymer molecules trapped between the graphene sheets effectively blocks all molecular and ionic permeation through the extra space of 0.6 Å in the composite membranes,

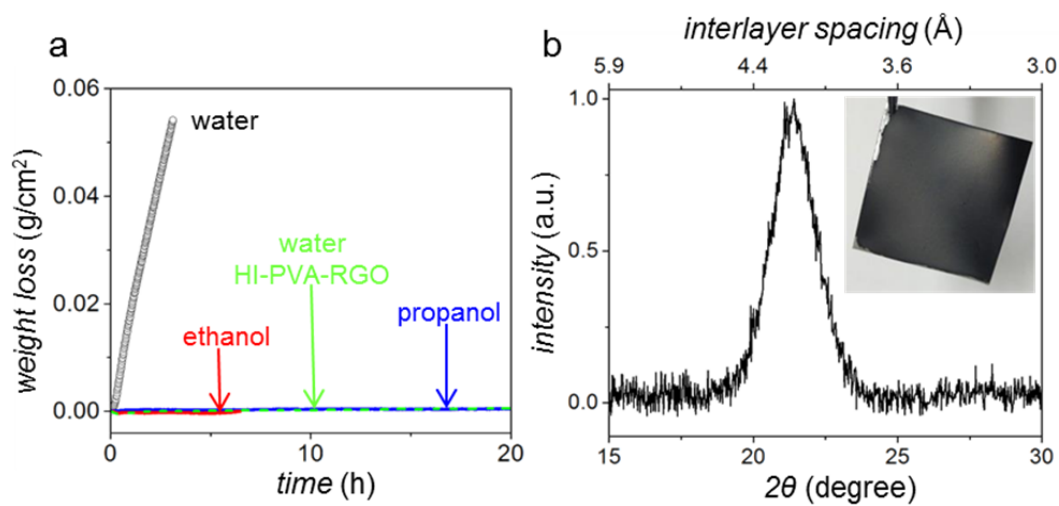

*Figure S3*. PVA-GO composites films exhibit barrier properties similar to those of GO laminates but with improved mechanical strength. *a* – Weight loss for a container filled with water or other liquids and sealed with a 1 μm thick PVA-GO membrane before and after its reduction with HI acid. The measurements were carried out at room temperature in a glove box (ref. 5 of the main text). The green curve shows water permeation after the reduction in HI; the other curves are for non-reduced PVA-GO. *b* – X-ray diffraction for HI-reduced PVA-GO membrane. Inset: Photograph of a 2 cm×2 cm steel plate coated with VC-reduced PVA-GO.

**S4. Supporting References**

1  Li, Y.-Q., Yu, T., Yang, T.-Y., Zheng, L.-X. & Liao, K. Bio-Inspired Nacre-like Composite Films Based on Graphene with Superior Mechanical, Electrical, and Biocompatible Properties. *Adv. mater.* **24**, 3426-3431 (2012).
2  Tian, Y., Cao, Y., Wang, Y., Yang, W. & Feng, J. Realizing Ultrahigh Modulus and High Strength of Macroscopic Graphene Oxide Papers Through Crosslinking of Mussel-Inspired Polymers. *Adv. mater.* **25**, 2980-2983 (2013).
3  An, Z., Compton, O. C., Putz, K. W., Brinson, L. C. & Nguyen, S. T. Bio-Inspired Borate Cross-Linking in Ultra-Stiff Graphene Oxide Thin Films. *Adv. mater.* **23**, 3842-3846 (2011).